\documentclass[journal]{IEEEtran}
\IEEEoverridecommandlockouts

\usepackage{tikz}
\usepackage{textcomp}
\usepackage{hyperref}
\usepackage{lipsum}

\newcommand\copyrighttext{%
  \footnotesize \textcopyright 2021 IEEE. Personal use of this material is permitted.
  Permission from IEEE must be obtained for all other uses, in any current or future 
  media, including reprinting/republishing this material for advertising or promotional 
  purposes, creating new collective works, for resale or redistribution to servers or 
  lists, or reuse of any copyrighted component of this work in other works. 
  DOI: 10.1109/LWC.2021.3095066}
\newcommand\copyrightnotice{%
\begin{tikzpicture}[remember picture,overlay]
\node[anchor=south,yshift=8pt] at (current page.south) {\fbox{\parbox{\dimexpr\textwidth-\fboxsep-\fboxrule\relax}{\copyrighttext}}};
\end{tikzpicture}%
}

\usepackage{amsthm,amsmath,amssymb}
\usepackage{mathrsfs}
\usepackage{color}
\usepackage{graphicx} 

\begin{document}
%
\title{ LMMSE-based SIMO Receiver for Ultraviolet Scattering Communication with Nonlinear Conversion }
%
%
%

\author{Zanqiu Shen,
        JianShe Ma,
        and Ping Su*
\thanks{The authors are with Division of Advanced Manufacturing, Tsinghua Shenzhen International Graduate School, Tsinghua University, Shenzhen 518055, China (e-mail: su.ping@sz.tsinghua.edu.cn, shen-zq18@mails.tsinghua.edu.cn). Zanqiu is also with Department of Precision Instrument, Tsinghua University, Beijing 100084, China.}
\thanks{}}

%
%

\markboth{}%
{Shell \MakeLowercase{\textit{et al.}}: Bare Demo of IEEEtran.cls for IEEE Journals}
%



\maketitle

\copyrightnotice

\begin{abstract}
Linear minimum mean square error (LMMSE) receivers are often applied in practical communication scenarios for single-input-multiple-output (SIMO) systems owing to their low computational complexity and competitive performance. However, their performance is only the best among all the linear receivers, as they minimize the bit mean square error (MSE) alone in linear space. To overcome this limitation, in this study, we propose an LMMSE receiver based on the measurements augmented by their nonlinear conversion for a photon-counting receiver, a photomultiplier tube, and an avalanche photodetector. The performance of the proposed LMMSE receiver is studied for different nonlinear conversions, numbers of receivers, and receiver types. The simulation results indicate that the Monte Carlo results are consistent with the analytical results and that the proposed LMMSE receiver outperforms the conventional one in terms of bit MSE and bit error rate. Accordingly, it can be concluded that to achieve a desired bit MSE, the proposed LMMSE-based nonlinear receiver not only reduces the need to increase the number of receivers but also reduces the bandwidth requirements.

\end{abstract}

\begin{IEEEkeywords}
Single-input-multiple-output (SIMO), Optical wireless scattering communication, Linear minimum mean square error (LMMSE), Nonlinear conversion.
\end{IEEEkeywords}

%
\IEEEpeerreviewmaketitle

\section{Introduction}
%
%
%
%
\IEEEPARstart{N}{on-line-of-sight} optical wireless scattering communication (OWSC) serves as a promising alternative to optical wireless communication, as it can maintain a certain information transmission rate even when the transmission link is blocked by obstacles or the transmitter and receiver are not perfectly aligned \cite{yuan2016review,vavoulas2019survey}. The typical OWSC system operates in the ultraviolet (UV) spectrum, which has strong scattering effects and enhances the received signals in a short range. However, the theoretical analyses \cite{shen2019modeling} as well as the experimental results \cite{wang20181mbps} of previous studies have revealed that the scattering and attenuation mechanisms lead to significant path loss and weak received signals. Therefore, the received signals cannot be detected by an optical waveform detector because they exhibit discrete photons. Thus, photon-level energy type detectors, such as photon counting (PC) receivers, photomultiplier tubes (PMTs), and avalanche photodetectors (APDs), are adopted for signal detection \cite{gong2016optical}. PC receivers detect the received signals by counting photons, whereas PMTs and APDs ﬁrst convert the received photons into electric current and then amplify the current to handle weak optical signal detection.

To improve the UV scattering communication performance, many technologies have been employed at both the transmitter and receiver ends. For example, several modulation techniques and channel coding schemes at the transmitter end have been proposed to address the limitation of significant path loss, including M-ary spectral amplitude code modulation \cite{noshad2013nlos}, nonuniform amplitude modulation \cite{gong2016analysis}, the low density parity check (LDPC) code \cite{qin2017received}, and the polar code \cite{hu2017research} scheme. At the receiver end, diversity techniques have often been employed to mitigate atmospheric turbulence effects. For example, Xiao \emph{et al.} proposed a maximum likelihood (ML) spatial diversity scheme for log-normal distributed atmospheric turbulence \cite{xiao2012bit}, and Arya \emph{et al.} proposed a dual-branch switch and stay diversity reception method assuming Gamma-Gamma distributed atmospheric fluctuations \cite{arya2018non}.  

Atmospheric turbulence in the transmission channel deteriorate the communication performance, and the random characteristics of the weak received optical signals can hinder the receiver signal processing. The received signals by UV scattering communications are modeled as a Poisson distribution for PC receivers and Poisson--Gaussian distribution for PMTs and APDs \cite{he2010performance}. In this case, the statistical properties of the output electrical signals for PMTs or APDs lead to high computational complexity in ML detection \cite{gong2015lmmse}. Therefore, Gong \emph{et al.} proposed a linear minimum mean square error (LMMSE) receiver that requires only the first and second order moments of the received signals alone \cite{gong2015lmmse}, \cite{liu2019power}. This LMMSE receiver is characterized by a lower computational complexity and competitive performance.

In this letter, we propose an LMMSE-based single-input-multiple-output (SIMO) receiver based on the augmentation of the received signals by its nonlinear transformations. The nonlinear conversions are performed using power functions with an arbitrary power. Then, we derive the statistical results of the augmented signals, which are related to the arbitrary order of moment of the augmented signals from the PC receiver, PMT, and APD. Based on these statistical results, we obtain the linear coefficients of the augmented measurements and the corresponding mean square error (MSE). Compared with \cite{gong2015lmmse}, the proposed receiver reduces the need to increase the number of receivers and also reduces the bandwidth requirements by OOK modulation via nonlinear conversion. Furthermore, the proposed receiver offers large room for improvement for the design of better nonlinear functions based on power functions. Thus, the proposed receiver is a generalization of the LMMSE receiver in \cite{gong2015lmmse} and presents a new route to construct a nonlinear receiver.

\section{System Model}
For on-off-keying (OOK) modulation, the transmit symbol is ``0" or ``1" with probability $P(B=1)=P(B=0)=\frac{1}{2}$. For M-pulse position modulation, the probability of the transmit pulse ``0" and ``1" is $P(B=0) = \frac{M-1}{M}$ and $P(B=1) = \frac{1}{M}$, respectively. We assume $K$ receivers at the receiver end, and the corresponding received signals are $z_1, z_2,..., z_K$. They can be combined linearly according to the LMMSE framework\cite{gong2015lmmse}, which is the best among all linear estimators \cite{bar2004estimation}. Thus, it can be improved only if the nonlinear estimators are considered, which can be realized by nonlinear conversions \cite{lan2015nonlinear}.

In this letter, we adopt power functions to perform the nonlinear transformation, $y_{n,i} = g(z_i) = z_i^{n}, \label{Nonlinear conversion}$ define $n$ as a nonlinear factor, and denote $y_n = [y_{n,1}, y_{n,2}, ..., y_{n,K}]^T$. 

\emph{Remark 1}: 
Under power function conversion, the statistical results for converted signals needed to construct the proposed nonlinear receiver are directly related to the moment information of the received signals, which can be easily obtained either analytically (from the analytical distribution) or numerically (from the received signal data). Moreover, the power functions can be viewed as building blocks of nonlinear functions because they can be approximated by the power series. The proposed receiver therefore offers large room for improvement for the design of better nonlinear functions based on power functions.

Augmenting the original received signals with nonlinear converted signals, we obtain the total received signals $x = [y_m^T\,y_n^T]^T$. The symbol estimation $\hat{A}$ is defined by \cite{lan2015nonlinear}
\begin{equation}
\hat{A} = \alpha_x^T x + b',
\end{equation}
where 
\begin{equation}
\setlength{\abovedisplayskip}{0pt}
\begin{split}
\alpha_x^T =& P_{Bx} P_{xx}^{-1},\\
b' =& \frac{1}{M}-\alpha_x^T \mathbb{E}[x],
\label{alpha_z, alpha_y, b}
\end{split}
\setlength{\abovedisplayskip}{0pt}
\end{equation}
where $P_{Bx} = [P_{By_m},P_{By_n}]$, $P_{y_m\,y_n} = P_{y_n y_m} = \mathbb{E}[(y_m-\mathbb{E}[y_m])(y_n-\mathbb{E}[y_n])^T]$, and $P_{xx} = \left[ \begin{matrix}  P_{y_m}& P_{y_m y_n}\\ P_{y_n y_m}& P_{y_n} \end{matrix} \right]$. The MSE of the estimator (\ref{alpha_z, alpha_y, b}) is \cite{lan2015nonlinear}
\begin{equation}
\setlength{\abovedisplayskip}{0pt}
\begin{split}
D = P_B-P_{By_m}P_{y_m}^{-1}P_{By_m}^{T} - (P_{By_n}-P_{By_m}P_{y_m}^{-1} P_{y_m y_n})\\
( P_{y_n}-P_{y_n y_m} P_{y_m y_m}^{-1}P_{y_m y_n}  )^{-1} (P_{By_n} - P_{B y_m}P_{y_m}^{-1} P_{y_m y_n})^{T}.  
\end{split}
\setlength{\abovedisplayskip}{0pt}
\label{D: LMMSE with nonlinear conversion}
\end{equation}
To develop the LMMSE-based nonlinear receiver with nonlinear conversion, we need to compute $P_B, P_{By_m}, P_{y_m}, P_{By_n}, P_{y_n}$, and $P_{y_m y_n}$, where $P_B = 1/M - 1/M^2$.

\section{Nonlinear Receiver with Nonlinear Conversion}
\subsection{Photon counting receiver}
For a PC receiver, the received signal $z_i$ follows the Poisson distribution $\mathbb{F}(z_i; \lambda) = \frac{\lambda^{z_i}}{z_i !} {\rm exp}[-\lambda],
\label{Poisson distribution}$ where $\lambda = \lambda_i+\lambda_b$ when $B=1$, $\lambda = \lambda_b$ when $B=0$, $\lambda_b$ is the average number of received photons due to background radiation noise, and $\lambda_i$ is the average number of received photons at the $i_{\rm th}$ receiver. Next, we need to determine the $k_{\rm th}$ order moment of $z_i$
\begin{equation}
\setlength{\abovedisplayskip}{0pt}
\mathbb{E}(z_i^{k}) = \sum_{l=0}^{k} s(k,l) \lambda^{l},
\label{kth moment of Poisson distribution}
\setlength{\belowdisplayskip}{0pt}
\end{equation}
where $s(k,l)$ represent the Stirling numbers of the second type.

\emph{Theorem 3.1: For $B=1$, we have}
\begin{flalign}
\mathbb{E}[y_{n,i} \mid B=1] 
= &\sum_{l=0}^{n} s(n,l) (\lambda_i+\lambda_b)^{l}, \label{eq:5}\\
\mathbb{E}[y_{n,i}^2 \mid B=1] 
= &\sum_{l=0}^{2n} s(2n,l) (\lambda_i+\lambda_b)^{l},\label{eq:6}\\
\mathbb{E}[y_{n,i} y_{n,j} \mid B=1] 
= &\sum_{k=0}^{m} \sum_{l=0}^{n} s(m,k) s(n,l) 
(\lambda_i+\lambda_b)^k \nonumber \\ 
&(\lambda_j+\lambda_b)^{l}.\label{eq:7}
\end{flalign}
\emph{For $B=0$, similar results can be obtained by setting $\lambda_i=\lambda_j=0$ in the results for $B=1$.}

\emph{Proof:}
{$\mathbb{E}(y_{n,i} \mid B = 1)$ is obtained by Eq. (\ref{kth moment of Poisson distribution}) and $\mathbb{F}(z_i;\lambda_i+\lambda_b)$ by substituting $y_{n,i}$ with $z_i^n$. $\mathbb{E}(y_{n,i}^2 \mid B = 1)$ is obtained by Eq. (\ref{kth moment of Poisson distribution}) and $\mathbb{F}(z_i;\lambda_i+\lambda_b)$ by substituting $y_{n,i}$ with $z_i^{2 n} $. For the $\mathbb{E}(y_{n,i} y_{n,j} \mid B = 1)$, we have $\mathbb{E}[y_{n,i} y_{n,j} \mid B = 1] = \mathbb{E}[y_{n,i} \mid B = 1] \mathbb{E}[y_{n,j} \mid B = 1]$ since the received signals at the $i_{\rm th}$ receiver and the $j_{\rm th}$ receiver are assumed to be conditionally independent, where $\mathbb{E}[y_{n,i} \mid B = 1] $ and $\mathbb{E}[y_{n,j} \mid B = 1]$ can be calculated by Eq. (\ref{eq:5}) .}

\cite[Theorem 1]{gong2015lmmse} is a special case of Theorem 3.1 in this letter by setting $n=1$. Using Theorem 3.1, the expectation value of the converted signal $y_{n,i}$ is derived as
\begin{equation}
\setlength{\abovedisplayskip}{0pt}
\mathbb{E}[y_{n,i}] = \frac{1}{M} \sum_{l=0}^{n} f(n,l,\lambda_i),
\label{Eyni}
\setlength{\belowdisplayskip}{0pt}
\end{equation}
where $f(x,y,z) = s(x,y)[(\lambda_b+z)^y + (M-1)\lambda_b^y]
\label{}$. 

\emph{Proof:}
{Using the law of total probability, we  have $\mathbb{E}(y_{n,i}) = \mathbb{E}[z_i^{n} \mid B = 1] P(B = 1) + E[z_i^{n} \mid B = 0] P(B = 0)$, where $\mathbb{E}[z_i^{n} \mid B = 1]$ and $\mathbb{E}[z_i^{n} \mid B = 0]$ can be obtained by Theorem 3.1.}

The cross-covariance of $B$ and $y_{n}$ for the PC receiver is
\begin{equation}
P_{By_n} = {\rm R}_{By_n} - {1}/{M} \mathbb{E}[y_{n}],
\label{PByn}
\end{equation}
where ${R}_{B y_{n,i}} = {1}/{M} \sum_{l=0}^{n} s(n,l)(\lambda_b+\lambda_i)^l \label{RByni}.$

\emph{Proof:}
{Using the law of total probability, we have $R_{B y_{n,i}} = \mathbb{E}[B y_{n,i}] = \mathbb{E}[B y_{n,i} \mid B=1] P(B=1) + \mathbb{E}[B y_{n,i} \mid B=0] P(B=0) = \mathbb{E}[y_{n,i} \mid B=1] P(B=1)$. We then obtain $\mathbb{E}[y_{n,i} \mid B = 1]$ from Theorem 3.1.
} 

The covariance matrix of $P_{y_n}$ is the difference of the correlation matrix ${R}_{y_{n}}$ and the multiplication of the expectation $1/M \mathbb{E}[{y_n}]$, which is
\begin{equation}
P_{y_n} = {R}_{y_n} - \mathbb{E}[y_n] \mathbb{E}[y_n]^T,
\label{Pyn}
\end{equation} and the  diagonal elements ${ R}_{y_{n,i} y_{n,i}} = {1}/{M} \sum_{l=0}^{2n} f(2n,l,\lambda_i)$ of the correlation matrix ${ R}_{y_n}$ with the off-diagonal elements ${ R}_{y_{n,i} y_{n,j}} = {1}/{M} \sum_{k=0}^{n} \sum_{l=0}^{n} g(n, n, k, l, \lambda_i, \lambda_j)$ of matrix ${R}_{y_n}$, where $g(u, v, w, x, y, z) = s(u, w) s(v, x) \lambda_b^{w+x}
[ (1+{y}/{\lambda_b})^w (1+{z}/{\lambda_b})^x + M-1 ].\label{f(x,y)}$

\emph{Proof:}
{Using the law of total probability, we have ${ R}_{y_{n,i} y_{n,i}} = \mathbb{E}[{y_{n,i}^{2}}] = \mathbb{E}[{z_i^{2 n}}]
= \mathbb{E}[z_i^{2 n} \mid B = 1] P(B = 1) + \mathbb{E}[z_i^{2 n} \mid B = 0] P(B = 0), $ which can be computed from Theorem 3.1. Using the law of total probability and the assumption of conditional independence \cite{gong2015lmmse}, we have ${R}_{y_{n,i} y_{n,j}} = \mathbb{E}[y_{n,i} y_{n,j}] 
= \mathbb{E}[{y_{n,i} y_{n,j}} \mid B = 1] P(B=1)+\mathbb{E}[{y_{n,i} y_{n,j}} \mid B = 0] P(B=0) = \mathbb{E}[{y_{n,i} \mid B = 1}] \mathbb{E}[{y_{n,j}} \mid B = 1] P(B=1)+\mathbb{E}[{y_{n,i}} \mid B = 0] \mathbb{E}[{y_{n,j}} \mid B = 0] P(B=0)$, which can be computed from Theorem 3.1.
}

Because the converted signals are not guaranteed to be uncorrelated with each other, the covariance matrix between $y_m$ and $y_n$ is not zero \cite{lan2015nonlinear}, which is defined as
\begin{equation}
P_{y_m y_n} = R_{y_m y_n} - \mathbb{E}[y_m], \mathbb{E}[y_n]^T,
\label{Pymyn}
\end{equation}
where ${R}_{y_{m,i} y_{n,i}} = {1}/{M} \sum_{l=0}^{m+n} f(m+n,l,\lambda_i)$, $
{\rm R}_{y_{m,i} y_{n,j}} = {1}/{M} \sum_{k=0}^{m} \sum_{l=0}^{n} g(m, n, k, l, \lambda_i, \lambda_j) \label{Rymiynj}$.

\emph{Proof:}
{Using the law of total probability, we get ${R}_{y_{m,i} y_{n,i}} = \mathbb{E}[{y_{m,i} y_{n,i}}] = \mathbb{E}[{z_i^{m+n}}]
= \mathbb{E}[z_i^{m+n} \mid B = 1] P(B = 1) + \mathbb{E}[z_i^{m+n} \mid B = 0] P(B = 0)$. Using the law of total probability and the assumption of conditional independence, we obtain ${R}_{y_{m,i} y_{n,j}} = \mathbb{E}[y_{m,i} y_{n,j}] 
= \mathbb{E}[{y_{m,i} y_{n,j}} \mid B = 1] P(B=1)+\mathbb{E}[{y_{m,i} y_{n,j}} \mid B = 0] P(B=0) = \mathbb{E}[{y_{m,i} \mid B = 1}] \mathbb{E}[{y_{n,j}} \mid B = 1] P(B=1)+\mathbb{E}[{y_{m,i}} \mid B = 0] \mathbb{E}[{y_{n,j}} \mid B = 0] P(B=0)$, which can be calculated by Theorem 3.1.
}

By substituting Eq. (\ref{Eyni})-(\ref{Pymyn}) into Eq. (\ref{alpha_z, alpha_y, b}), we obtain the LMMSE receiver with nonlinear conversion for the PC receiver. The corresponding MSE is obtained from Eq. (\ref{D: LMMSE with nonlinear conversion}).

\subsection{PMT or APD}
For a PMT or APD receiver, the received signal $z_i$ follows $\mathbb{P}(z_i;\lambda) = \sum_{n_i=0}^{+\infty} \frac{\lambda_i^{n_i}}{n_i!} \exp[-\lambda_i] \cdot \mathcal{N}(z_i;n_i A e, n_i \sigma^2 + \sigma_0^2),\label{PMT APD zi}$, where $\lambda = \lambda_i+\lambda_b$ when $B=1$ and $\lambda = \lambda_b$ when $B=0$ \cite{he2010performance}, $\mathcal{N}(x;\mu,\sigma^2)$ denotes the Gaussian distribution with mean $\mu$ and variance $\sigma^2$, $A$ is the amplification factor, $e$ is the single electron charge, $\sigma^2$ denotes the shot noise variance stimulated by a single photon, and $\sigma_0^2$ denotes the thermal noise variance \cite{gong2015lmmse}. Next, we need to determine the $k_{th}$ order moment of the Gaussian distribution, which is
\begin{equation}
\mathbb{E}[z_i^n] = \sum_{k=0}^{\lfloor n/2 \rfloor} C_n^{2 k} \mu^{n-2k} \frac{(2k)!}{k! 2^k}\sigma^{2k}.
\label{eq:12}
\end{equation}
\emph{Theorem 3.2: For $B=1$, we have}
\begin{flalign}
\mathbb{E}[y_{n,i} \mid B=1]=
& \sum_{k=0}^{\lfloor n/2 \rfloor} \sum_{l=0}^{k} \sum_{m=0}^{n-k-l} I(n,k,l) s(n-k & \nonumber\\ 
& -l,m) (\lambda_i+\lambda_b)^m, \\
\mathbb{E}[y_{n,i}^2 \mid B=1]  =
& \sum_{k=0}^{ n } \sum_{l=0}^{k} \sum_{m=0}^{2n-k-l} I(2n,k,l) s(2n- & \nonumber\\ 
& k-l,m) (\lambda_i+\lambda_b)^m, \\
\mathbb{E}[y_{n,i} y_{n,j} \mid B=1] =
& \sum_{k_1 = 0}^{\lfloor n/2 \rfloor} \sum_{l_1 =0}^{ k_1} \sum_{m_1=0}^{n-k_1-l_1} \sum_{k_2 = 0}^{\lfloor n/2 \rfloor} \sum_{l_2 =0}^{k_2} 
\sum_{m_2=0}^{n-k_2-l_2}  &\nonumber\\
&[I(n,k_1,l_1)
I(n, k_2, l_2) s(n-k_1-l_1, &\nonumber\\
& m_1) s(n-k_2-l_2, m_2) (\lambda_i+\lambda_b)^{m_1} &\nonumber\\
&(\lambda_j+\lambda_b)^{m_2},
\label{thm2}
\end{flalign}
where $I(x,y,z) = C_{x}^{2 y} (A e)^{x-2 y} \sigma^{2(y-z) \sigma_{0}^{2 z} \frac{(2 y )!}{y! 2^k}}. \label{I(x,y,z)}$
\emph{For $B=0$, similar results can be obtained by setting $\lambda_i=\lambda_j=0$ in the results for $B=1$.}

\emph{Proof:}
\begin{flalign}		
\setlength{\abovedisplayskip}{0pt}
\mathbb{E}[y_{n,i} \mid B=1] 
= & \int_{-\infty}^{+\infty} z_i^n \mathbb{P}(z_i;\lambda_b+\lambda_i) {\rm d} z_i \label{eq:16}\\
=& \sum_{n=0}^{+\infty} \frac{(\lambda_i+\lambda_b)^n}{n!} \exp[-(\lambda_i+\lambda_b)] \sum_{k=0}^{\lfloor n/2 \rfloor}
&\nonumber\\
& C_n^{2 k} (n_i A e)^{n-2k} \frac{(2k)!}{k! 2^k} (n_i \sigma^2 + \sigma_0^2)^{k} \label{eq:17}\\
=& \sum_{n_i=0}^{+\infty} \sum_{k=0}^{\lfloor n/2 \rfloor} \sum_{l=0}^{k} I(n,k,l) n_i^{n-k-l}  \frac{(\lambda_i+\lambda_b)^n_i}{n_i!}
&\nonumber\\
&\exp[-(\lambda_i+\lambda_b)] \label{eq:18}\\
=& \sum_{k=0}^{\lfloor n/2 \rfloor} \sum_{l=0}^{k} \sum_{m=0}^{n-k-l} I(n,k,l) s(n-k-l,m)
&\nonumber\\ 
& (\lambda_i+\lambda_b)^m. \label{eq:19}
\setlength{\abovedisplayskip}{0pt}
\end{flalign}

{According to the definition of conditional expectation, we have Eq. (\ref{eq:16}). Using Eq. (\ref{eq:12}), we obtain Eq. (\ref{eq:17}). Grouping like terms together in Eq. (\ref{eq:17}), we then have Eq. (\ref{eq:18}). Using Eq. (\ref{kth moment of Poisson distribution}) to express the $(n-k-l)_{\rm th}$ order moment of Poisson distribution, we obtain Eq. (\ref{eq:19}). The derivation of $\mathbb{E}[y_{n,i}^2 \mid B = 1]$ is similar; therefore, we have omitted it. $\mathbb{E}[y_{n,i} y_{n,j} \mid B = 1]$ can be obtained by the assumption of conditional independence; therefore, we have omitted it.
}

Theorem 3.2 is a generalization of \cite[Theorem 4]{gong2015lmmse}. According to Theorem 3.2, we derive the following statistical results for the PMT or APD. The derivation processes for PMTs and APDs are similar to that of the PC receiver; hence, we directly present the results. The expectation value of $y_{n,i}$ is
\begin{equation}
\setlength{\abovedisplayskip}{0pt}
\mathbb{E}[y_{n,i}] = \frac{1}{M} \sum_{k=0}^{\lfloor n/2 \rfloor} \sum_{l=0}^{k} \sum_{m=0}^{n-k-l} I(n,k,l) f(n-k-l,m,\lambda_i).
\label{Eyni for PMT APD}
\setlength{\abovedisplayskip}{0pt}
\end{equation}
The cross-covariance of $B$ and $y_{n}$ for PMT or APD is represented by Eq. (\ref{PByn}), where
\begin{align}
\setlength{\abovedisplayskip}{1pt}
R_{By_{n,i}} = &\frac{1}{M} \sum_{k = 0}^{\lfloor n/2 \rfloor} \sum_{l =0}^{k} \sum_{m=0}^{n-k-l} I(n,k,l) s(n-k-l, m) \nonumber\\
&(\lambda_i+\lambda_b)^m. 
\label{}
\setlength{\abovedisplayskip}{0pt}
\end{align}
The covariance matrix of $y_n$ for PMT or APD is represented by Eq. (\ref{Pyn}), where
\begin{align}
\setlength{\abovedisplayskip}{0pt}
R_{y_{n,i} y_{n,i}} =& \frac{1}{M} \sum_{k = 0}^{n} \sum_{l =0}^{k} \sum_{m=0}^{2n-k-l} I(2n,k,l) \nonumber\\
                     &f(2n-k-l,m,\lambda_i) ,\\
R_{y_{n,i} y_{n,j}} =& \frac{1}{M} \sum_{k_1 = 0}^{\lfloor n/2 \rfloor} \sum_{l_1 =0}^{k_1} \sum_{m_1=0}^{n-k_1-l_1} \sum_{k_2 = 0}^{\lfloor n/2 \rfloor} \sum_{l_2 =0}^{k_2} \sum_{m_2=0}^{n-k_2-l_2} [I(n, k_1,\nonumber\\
                     &l_1)
I(n, k_2, l_2) g(n-k_1-l_1, n-k_2-l_2, m_1, \nonumber\\
                     &m_2, \lambda_i, \lambda_j)].
\label{Ryniynj for PMT APD}
\setlength{\belowdisplayskip}{0pt}
\end{align}

The covariance matrix of $y_m$ and $y_n$ for PMT or APD is represented by Eq. (\ref{Pymyn}), where
\begin{align}
\setlength{\abovedisplayskip}{0pt}
R_{y_{m,i} y_{n,i}} =& \frac{1}{M} \sum_{k = 0}^{\lfloor \frac{m+n}{2} \rfloor} \sum_{l =0}^{k} \sum_{p=0}^{m+n-k-l} [I(m+n,k,l) \nonumber \\
                     &f(m+n-k-l,p,\lambda_i)],\\
R_{y_{m,i} y_{n,j}} =& \frac{1}{M} \sum_{k_1 = 0}^{\lfloor m/2 \rfloor} \sum_{l_1 =0}^{k_1} \sum_{p_1=0}^{m-k_1-l_1} \sum_{k_2 = 0}^{\lfloor n/2 \rfloor} \sum_{l_2 =0}^{k_2} \sum_{p_2=0}^{n-k_2-l_2} [I(m,k_1,\nonumber\\ 
                     &l_1)
I(n, k_2, l_2) g(m-k_1-l_1, n-k_2-l_2, \nonumber\\
                     &p_1, p_2, \lambda_i, \lambda_j)].
\label{Rymiynj for PMT APD}
\setlength{\belowdisplayskip}{0pt}
\end{align}
By substituting Eq. (\ref{Eyni for PMT APD})-(\ref{Rymiynj for PMT APD}) into Eq. (\ref{alpha_z, alpha_y, b}) and (\ref{D: LMMSE with nonlinear conversion}), we obtain the LMMSE receiver with arbitrary power function conversions for PMT or APD and the corresponding bit estimation MSE. 

\emph{Remark 2}: 
{We need to compute $P_{y_n}^{-1}$ for the LMMSE receiver, and compute $P_{y_n}^{-1}$ and  $P_{y_m y_n}^{-1}$ for the LMMSE-based nonlinear receiver. The computation complexity is $\Theta((S+1)^3)$ for computing inverse matrix, where $S$ is the order of the matrix \cite{gong2015lmmse}. Therefore, the proposed receiver requires twice the computation time of the LMMSE receiver.}

\section{Numerical Results}
We compare the bit MSEs of the LMMSE receivers with and without nonlinear conversions, which are calculated by the LMMSE estimator in \cite{bar2004estimation} and Eq. (\ref{D: LMMSE with nonlinear conversion}). The simulation parameters were set as follows: receiver temperature $T^o=300~{\rm K}$, load resistance $R_L=5~{\rm M \Omega}$, APD ionization factor $\gamma=0.028$, PMT spreading factor $\xi=0.10$, quantum efficiency $\eta=0.06$, loss factor $L=4\times10^{10}$, transmission bit rate $R_b=1{\rm Mbps}$, Planck constant $h=6.62606957\times10^{-34}$, photon noise rate $\Lambda_b=20000~{\rm s^{-1}}$, Boltzmann constant $k_e=1.3806505\times 10^{-23}$, wavelength of transmitting light = $250~{\rm nm}$, and single electron charge $e=1.602\times10^{-19} {\rm C}$. We adopt PC receiver with OOK modulation and choose the nonlinear factor as $(m, n) = (1, 2)$ unless otherwise stated.

\begin{figure}[tb]
\vspace{-1.5cm} 
\setlength{\abovecaptionskip}{0cm}
\centering
\includegraphics[width=0.85\linewidth]{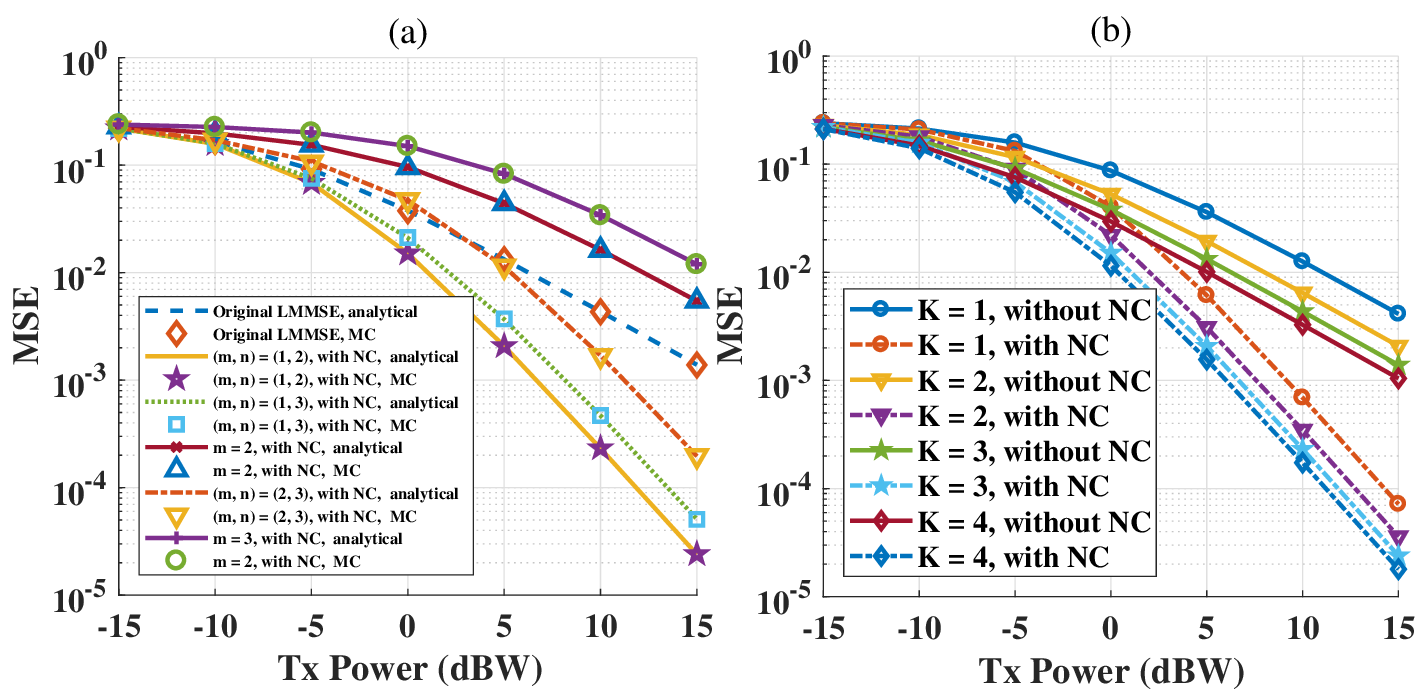}
\caption{(a) Analytical and Monte-Carlo (MC) MSE results from different nonlinear conversion (NC) combinations. (b) Analytical MSE results for PC receiver using different numbers of receivers with and without nonlinear conversions (NC).}
\label{fig:1}
\setlength{\belowcaptionskip}{-2cm}
\end{figure}

In Fig. \ref{fig:1}(a), we plot the MSE results for different nonlinear conversion combinations.
In Fig. \ref{fig:1}(a), we provide the Monte-Carlo (MC) results and the analytical results. The MC simulation process is as follows: simulate $10^5$ OOK symbols and sample the received numbers of photons of $K$ receivers for each OOK symbol; thereafter, use Eq. (\ref{alpha_z, alpha_y, b}) to combine the augmented signals to obtain the estimated symbol; then calculate $10^5$ square error values between the estimated symbols and the transmit symbols and, finally, average them to obtain the MSE results. From Fig. 1(a), we see that the MC results agree well with the analytical results. Moreover, fig. \ref{fig:1}(a) shows that the nonlinear conversion of $(m,n)=(1,2)$, $(1,3)$, and $(2,3)$ in the proposed receiver can outperform that in the conventional LMMSE receiver. However, the performance of the former for $m=2$ and $3$ is poorer than that of the latter. This implies that the performance gain through nonlinear conversion can only be obtained by appropriate nonlinear conversions. In Fig. \ref{fig:1}(b), we plot the MSE results for different numbers of receivers with and without nonlinear conversions. The MSE results are significantly reduced when nonlinear conversions are adopted in the LMMSE framework. The performance gain from the nonlinear conversions may exceed the performance gain from increasing the number of receivers. For example, when the transmitter power is $0~{\rm dBW}$, the MSE with nonlinear conversion at $K=2$ is 0.02199, which is smaller than that without nonlinear conversion at $K=4$, which is 0.02953. Therefore, we can use nonlinear conversion instead of increasing the number of receivers to achieve a desired MSE.

\begin{figure}[tb]
\vspace{-1.5cm} 
\setlength{\abovecaptionskip}{0cm}
\setlength{\belowcaptionskip}{-2.5cm}
\centering
\includegraphics[width=0.85\linewidth]{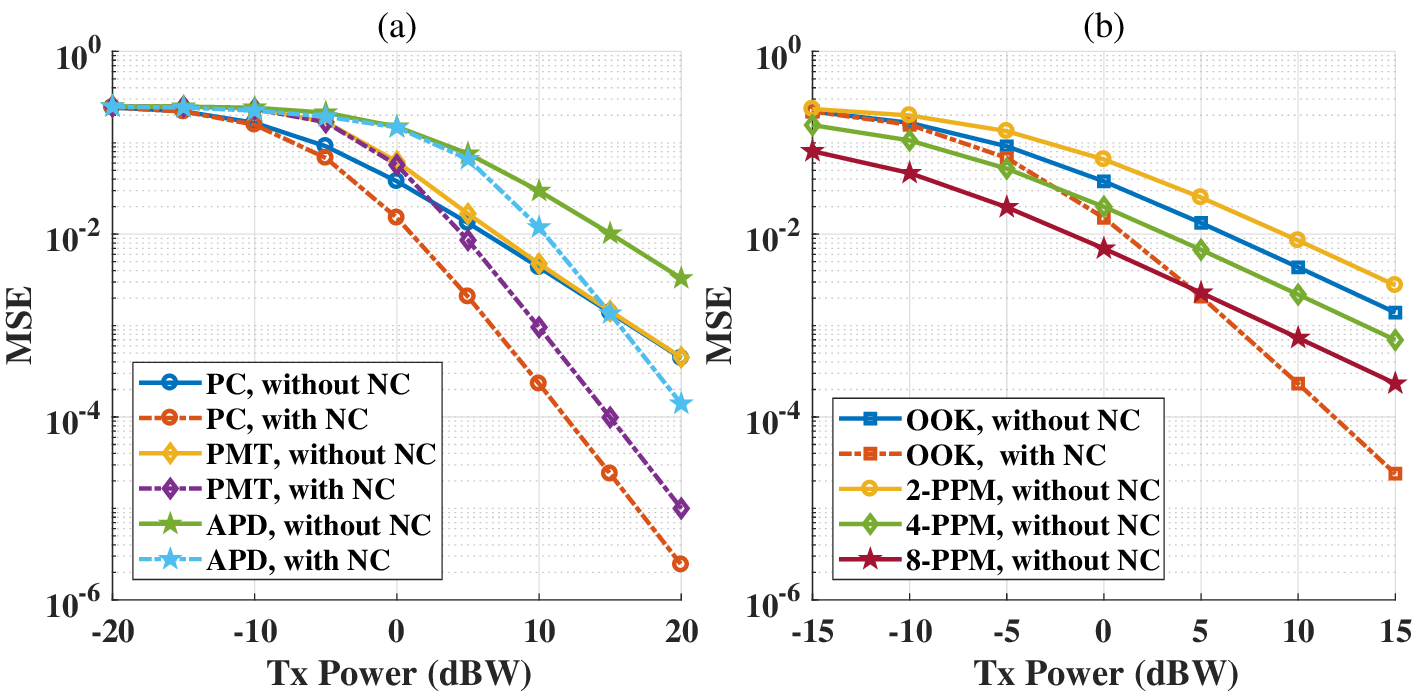}
\caption{(a) Analytical MSE results for different types of receivers with and without nonlinear conversions (NC). (b) Analytical MSE results for different modulation schemes with and without nonlinear conversions (NC).}
\label{fig:2}
\end{figure}

\begin{figure}[tb]
\vspace{-1cm} 
\setlength{\abovecaptionskip}{0cm}
\setlength{\belowcaptionskip}{-2cm}
\centering
\includegraphics[width=0.8\linewidth]{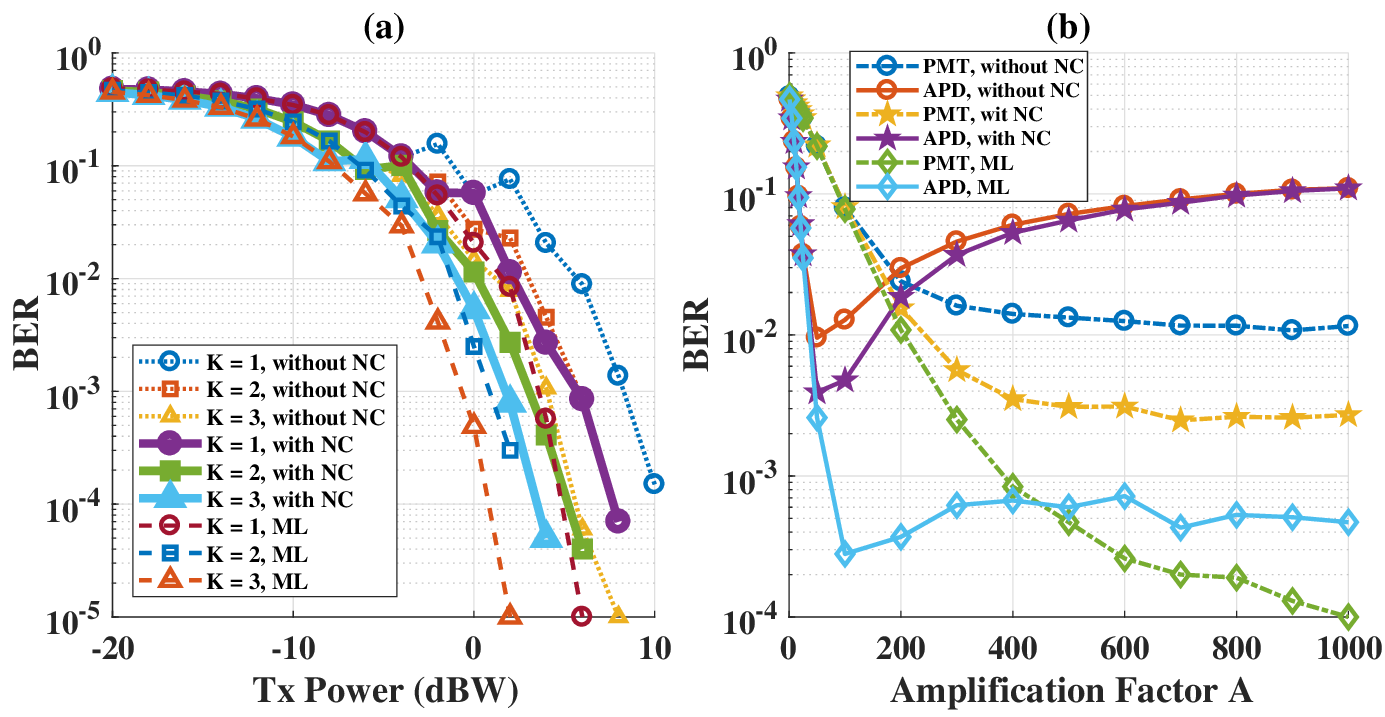}
\caption{(a) Simulated BER results for the Maximum likelihood (ML) receiver and the LMMSE receiver with and without nonlinear conversion. (b) Simulated BER results for PMT and APD receivers. The transmitter power is $8~{\rm dBW}$ for APD, and $1~{\rm dBW}$ for PMT. NC denotes nonlinear conversion.}
\label{fig:3}
\end{figure}

In Fig. \ref{fig:2}(a), we plot the MSE results from different types of receivers, including the PC receiver, PMT, and APD. The number of receivers is set as $K=3$. Fig. \ref{fig:2}(a) shows that the nonlinear conversions can reduce the MSE for different types of receivers. The performance gains of the LMMSE with nonlinear conversions become significant when the transmitter power exceeds $-10~{\rm dBW}$ for the PC receiver, $0~{\rm dBW}$ for the PMT, and 0~${\rm dBW}$ for the APD. The performance gains at $15~{\rm dBW}$ are 57.9 for the PC receiver, 14.5 for the PMT, and 7.5 for the APD. Moreover, the MSE of the APD with nonlinear conversions is smaller than that of the PC receiver or PMT without nonlinear conversions when the transmitter power $P_t$ is greater than $15~{\rm dBW}$, which indicates that APD may perform better at a large transmitter power. 
In Fig. \ref{fig:2}(b), we plot the MSE results with and without nonlinear conversions for different modulation schemes. From Fig. \ref{fig:2}(b), we observe that the MSE for the OOK modulation is smaller than that for the 2-${\rm PPM}$ modulation but larger than that for higher-order PPM modulation without nonlinear conversions when the transmitted power exceeds a certain value. For example, when the transmitted power is 10~${\rm dBW}$, the MSE is 0.000231 for the OOK modulation with nonlinear conversions, whereas it is 0.000732 for the 8-${\rm PPM}$ modulation without nonlinear conversions. This indicates that to achieve a desired bit MSE, we can utilize OOK modulation with nonlinear conversions to save bandwidth requirements. For example, assuming the desired MSE is $2 \times 10^{-3}$, the bandwidth requirement for OOK modulation and 8-PPM are $R_b$ and $\frac{8}{3} R_b$, respectively \cite{ghassemlooy2019optical}. We also obtain the MSE results with nonlinear conversions by 2-PPM, 4-PPM, and 8-PPM, which are less than the MSE results without nonlinear conversions. However, the BER performance is not improved correspondingly. Therefore, we only consider OOK modulation in Fig. \ref{fig:3}.

In Fig. \ref{fig:3}(a), we plot the BER results for the ML receiver and the LMMSE receiver with and without nonlinear conversions. The modulation scheme is OOK with a detection threshold fixed as 0.5.From Fig. \ref{fig:3}(a), we see that the LMMSE receiver with nonlinear conversions performs better than the conventional one but  worse than the ML receiver. For example, when $K=2$ and the transmitted power is $2~{\rm dBW}$, the BER with and without nonlinear conversion is 0.02286 and 0.00272, respectively, whereas the BER for the ML receiver is 0.0003. Fig \ref{fig:3}(b)  presents a plot of the BER results for the PMT and APD. We see that the proposed receiver performs better than the LMMSE receiver but worse than the ML receiver. {However, the ML receiver has a higher computation complexity since it typically needs to compute 50 terms in the nominator and denominator to estimate one symbol \cite{gong2015lmmse} while the proposed receiver only needs to do one linear combination.} We also observe that for the PMT, the proposed LMMSE receiver has a lower BER floor than that of the conventional one. Regarding the APD, the LMMSE receiver with nonlinear conversions outperforms the conventional one at only a certain range of amplification factors. For example, the BER for the LMMSE receiver with nonlinear conversions is less than that of the conventional one by 2.68 when $A$ is 100, whereas it is 1.14 when $A$ is 400.

\section{Conclusion}
In this letter, we proposed an LMMSE-based nonlinear receiver with nonlinear conversions for a PC receiver, PMT, and APD. The simulation results suggest that we can adopt a smaller number of receivers with nonlinear conversions instead of increasing the number of receivers without nonlinear conversions, thereby saving the receiver resources and simplifying its design. Moreover, we can select OOK modulation scheme with nonlinear conversions instead of higher-order modulation schemes without nonlinear conversions to reduce bandwidth requirements. In the future, a better designed nonlinear function is desired to improve the proposed receiver, with the aim of approaching the performance of the ML receiver.

The proposed receiver is generally applicable in all weak received optical signal scenarios where the received number of photons can be modeled as a Poisson distribution, such as long-distance or blocked visible-light communications. The proposed receiver can also be extended to MIMO systems \cite{wilson2005optical}, and the concept of nonlinear conversion can be used in pure wireless channels for symbol estimation or signal detection.

\section*{Acknowledgment}

The authors would like to thank Professor Chen Gong from University of Science and Technology of China for valuable discussions with him. We thank the reviewers for their valuable comments and suggestions.

\bibliographystyle{IEEEtran}
\bibliography{ref}

\end{document}